*Article*

# Self-Supervised Learning for Solar Radio Spectrum Classification

**Siqi Li** [1], **Guowu Yuan** [1,2,*], **Jian Chen** [1], **Chengming Tan** [3,4] **and Hao Zhou** [1]

[1] School of Information Science & Engineering, Yunnan University, Kunming 650500, China; 15926684519@mail.ynu.edu.cn (S.L.); jianchen@mail.ynu.edu.cn (J.C.); zhouhao@ynu.edu.cn (H.Z.)
[2] CAS Key Laboratory of Solar Activity, National Astronomical Observatories, Beijing 100012, China;
[3] State Key Laboratory of Space Weather, National Space Science Center, Beijing 100864, China; tanchengming@nssc.ac.cn
[4] School of Astronomy and Space Sciences, University of Chinese Academy of Sciences, Beijing 100049, China
* Correspondence: gwyuan@ynu.edu.cn

**Abstract:** Solar radio observation is an important way to study the Sun. Solar radio bursts contain important information about solar activity. Therefore, real-time automatic detection and classification of solar radio bursts are of great value for subsequent solar physics research and space weather warnings. Traditional image classification methods based on deep learning often require considerable training data. To address insufficient solar radio spectrum images, transfer learning is generally used. However, the large difference between natural images and solar spectrum images has a large impact on the transfer learning effect. In this paper, we propose a self-supervised learning method for solar radio spectrum classification. Our method uses self-supervised training with a self-masking approach in natural language processing. Self-supervised learning is more conducive to learning the essential information about images compared with supervised methods, and it is more suitable for transfer learning. First, the method pre-trains using a large amount of other existing data. Then, the trained model is fine-tuned on the solar radio spectrum dataset. Experiments show that the method achieves a classification accuracy similar to that of convolutional neural networks and Transformer networks with supervised training.

## 1. Introduction

Solar radio spectrum observations are an important tool for studying solar outbursts, which contain important information about solar activity [1]. The solar radio spectra are divided into various types, corresponding to different physical events [2,3]. With the development of radio spectrometers and the massive observational data trend, manual detection and classification of solar radio spectra can no longer meet the needs of research. Therefore, it is important to automatically detect and classify solar radio bursts from this massive information efficiently and rapidly for subsequent scientific research and space weather warning and forecasting.

The solar broadband radio spectrometers (SBRS) at the National Astronomical Observatory of the Chinese Academy of Sciences were put into operation during the 23rd solar activity cycle [4]. The devices have produced a large number of observations. However, since solar radio bursts are a low-probability event, the observed spectra of solar radio bursts are very small. Additionally, due to the presence of interference, the raw data are not clearly characterized and it is difficult to quickly distinguish between different kinds of data. This creates difficulties for subsequent astronomical studies [5]. Therefore, it is of great help for solar physics research to classify solar radio spectra accurately, quickly and automatically.

With the rapid development of hardware levels and artificial intelligence, an increasing number of deep learning models and algorithms are used to solve tasks related to natural language processing [6] and computer vision [7,8]. For astronomy problems, many previous works were carried out on solar radio spectrum classification and these works have used image processing algorithms, neural network models and deep learning [9]. P. J. Zhang et al. designed an event recognition analysis system that can automatically detect solar type III radio bursts. This system used Hough transform to recognize the line segment associated with type III bursts in the dynamic spectra [10]. However, the computational parameters of this method must be artificially designed and are not universally applicable. In recent years, with the development of convolutional neural networks (CNN) [11,12], long short-term memory (LSTM) networks [13], and deep confidence networks [14], many of these methods were applied to the classification of solar radio spectra. S. M. J. Jalali introduced LSTM [15], which was combined with a CNN to propose the CNN–LSTM approach [16]. This method improves performance with similar time consumption. B. Yan used a feature pyramid network (FPN) as a backbone network [17] and used
ResNet to extract features [18]. By simply connecting to the structure, FPN fuses features of different scales and different levels of semantics. The performance of detection is improved without affecting the speed of detection. In other related studies, a classification algorithm based on joint convolutional neural networks and transfer learning was proposed by using the inherent correlation between natural datasets and astronomical datasets. In addition, a cost-sensitive multiclassification loss function was proposed to make the network pay more attention to categories with fewer samples during the training process and use the meta-learning method to classify the solar radio spectrum with fewer samples. In addition, H. Salmane et al. proposed automatic identification methods for specific types of solar radio burst structures [19].

In current research on image classification, a large amount of training data is generally needed. However, since solar radio bursts are low-probability events, there are few samples. To solve the problem of fewer data, some studies have used transfer learning technology and improved loss functions. However, the large difference between natural images and solar radio spectrum images is not conducive to the application of transfer learning.

In natural language processing, self-supervised learning methods have become popular and these methods no longer require large amounts of labeled data in the training step. Researchers can design some rules to let the data supervise their own training [20]. For example, bidirectional encoder representations from Transformer (BERT) are designed as a kind of fill-in-the-blank method by masking some words in a sentence and then letting the network guess those words. In natural language processing, self-supervised learning is usually based on autoregressive language modeling in generative pre-training (GPT) and mask self-coding in BERT. Their basic principle is to delete some data and let the network learn to predict the deleted content.

In this paper, a solar radio spectrum classification method based on self-supervised learning is proposed. The method uses BERT to train the network to classify solar radio spectrum images by randomly masking a portion of the solar radio spectrum images and letting the network fill in the blank. The main contribution of this paper is that we apply self-supervised learning to classify the solar radio spectrum for the first time. This method is more conducive to using transfer learning. This paper can provide a reference for other small sample data classifications in astronomy.

## 2. Solar Radio Spectrum Dataset and its Preprocessing

The solar radio spectra are obtained from the solar broadband radio spectrometer (SBRS) at the National Astronomical Observatory of the Chinese Academy of Sciences. The raw data are stored in binary format and visualized to obtain the solar radio spectrum images. The vertical axis of the solar radio spectrum image represents the frequency of the spectrum, the horizontal axis represents the time, and each pixel value represents the radio flux of the Sun at a certain time and frequency. When displayed as a grayscale image, white indicates high solar radio flux, black indicates low solar radio flux, and the whole image represents the solar radio flux at a frequency over a period of time. The solar radio spectra are divided into three categories: burst, calibration and non-burst, as shown in Figure 1. The classification tasks in this paper are aimed at these three categories.

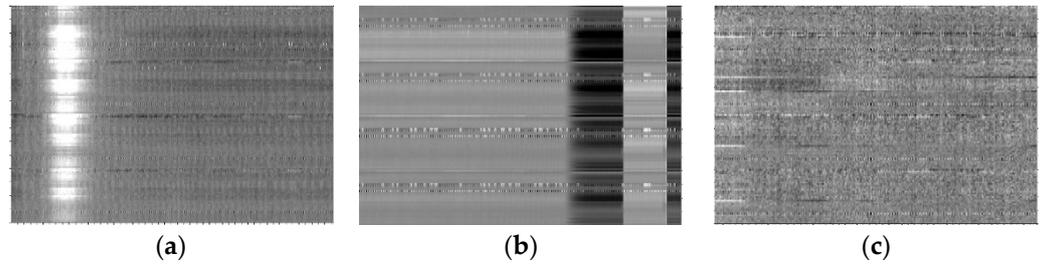

(a) (b) (c)

**Figure 1.** Solar radio spectrum data types. (**a**) Radio burst; (**b**) calibration; (**c**) non-burst.

The solar radio spectra images show that the original image is noisy and most of the noise is transverse stripe noise. This noise will affect the subsequent classification accuracy. The frequency channel normalization can reduce transverse stripe noise and make its features obvious, and the calculation method is as follows:

$$p'(x,y) = p(x,y) - \frac{1}{n}\sum_{y=0}^{n} p(x,y) + \frac{1}{mn}\sum_{x=0}^{m}\sum_{y=0}^{n} p(x,y) \qquad (1)$$

where $p(x,y)$ represents the radio intensity at time $x$ and frequency $y$ on the spectrum and $p'(x,y)$ is the radio intensity after channel normalization. The final result is shown in Figure 2. The noise of the horizontal stripe is significantly reduced and the burst is more obvious. It makes the features of the image more obvious and facilitates the learning of the subsequent network.

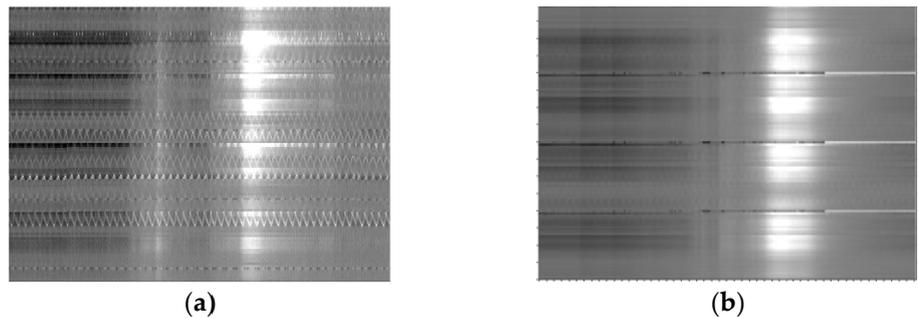

(a) (b)

**Figure 2.** Results of channel normalization. (**a**) Before channel normalization; (**b**) after channel normalization.

## 3. Method Section

### 3.1. Self-Supervised Learning

Self-supervised learning (SSL) is the main method used with the Transformer model to learn from large-scale unlabeled datasets [21]. The basic idea of SSL is to fill in the gaps. It masks or hides some parts of the input and uses observable parts to predict hidden parts [22]. Another effective method of self-supervised learning is contrastive learning [23]. In this case, we usually learn the feature representation of samples by creating positive and negative samples and comparing the data with positive samples and negative samples in the feature space. The advantage of contrastive learning is that it does not need to reconstruct pixel-level details to obtain image features but only needs to learn differentiation in the feature space. However, the construction of positive and negative samples is a difficult point in contrastive learning.

SSL provides a promising learning paradigm since it enables learning from a vast amount of readily available nonannotated data. SSL is implemented in two steps. First, a model is trained to learn a meaningful representation of the underlying data by solving a pretext task. The pseudo labels for the pretext task are automatically generated based on data attributes and task definition. Therefore, a critical choice in SSL is the definition of a pretext task. Second, the model trained in the first step is fine-tuned on the downstream tasks using labeled data. The downstream tasks include image classification and object detection [24,25].

*3.2. Self-Supervised Learning with Self-Masking*

Self-supervised learning methods are widely used in natural language processing. However, due to the large difference between the CNN model in computer vision and the Transformer used in natural language processing, the natural language processing method cannot be effectively transferred to computer vision tasks. However, when Vision Transformer (ViT) was proposed [26], the channel between computer vision and natural language was opened [27–30].

We refer to the mask method used in the BERT and GPT in natural language and let the model learn to restore sentences after covering some words. When the methods are transferred to solar radio spectrum classification, the solar radio spectrum image is first divided into blocks, which are equivalent to a word in a sentence. Then, these blocks are masked randomly and sent to the network. The network is composed of the encoder and decoder of ViT. After the image is restored through model learning, the trained encoder is removed and connected to the fully connected layer for classification. The structure of the self-mask model is shown in Figure 3.

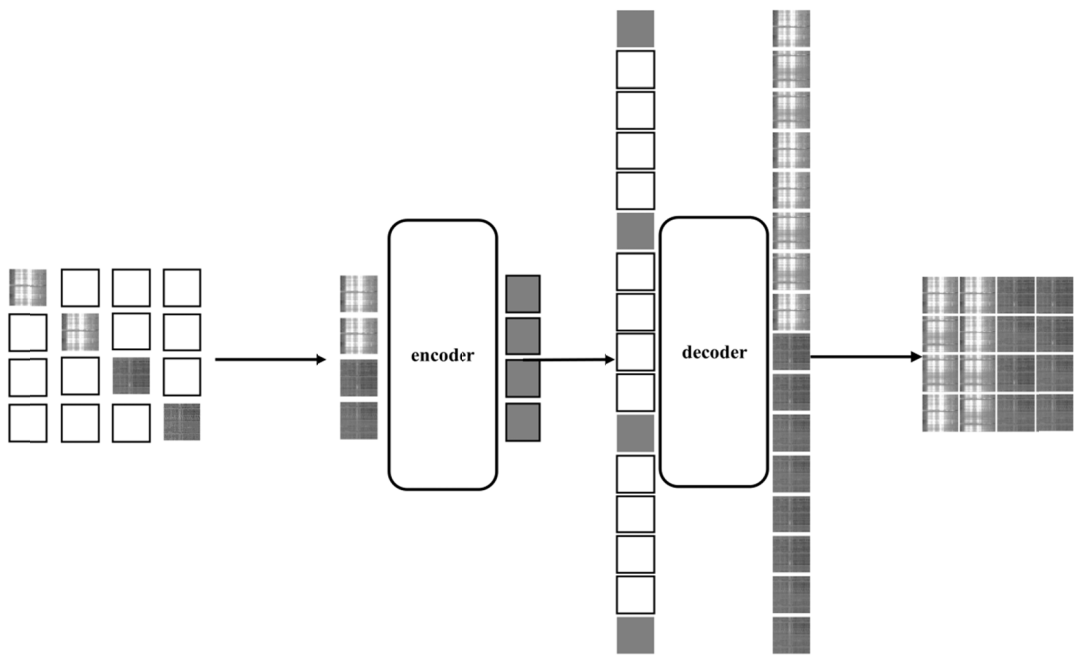

**Figure 3.** Self-masking model structure.

Compared with text, we can cover more parts due to the more redundant information in the image. This not only saves space and improves the model speed but also helps the model learn more information from solar radio spectrum images.

*3.3. Encoder and Decoder*

Our encoder is only derived from the ViT structure for visible unmasked blocks. Similarly to standard ViT, our encoder embeds patches by adding a linear projection of the positional embedding and then processes the dataset through a series of Transformer blocks. However, our encoder only needs to work on a small subset (e.g., 25%), which allows us to train very large encoders that require only a fraction of the computations and memory. The complete dataset is processed by a lightweight decoder.

The input to the decoder is a complete token set consisting of encoded visible blocks and mask tokens. Each mask token is a shared and learned vector that represents the missing blocks to be predicted. We add a position to all tokens in this complete set. If we do not do this, mask tokens will not have information about their positions in the image. The decoder also has another series of Transformer blocks.

The decoder is only used to perform image reconstruction tasks prior to training. Thus, the decoder architecture can be designed in a flexible way, independent of the encoder design. We experiment with very small decoders that are narrower and shallower than the encoder. For example, our default decoder is 10% smaller in computations per token than the encoder. In this asymmetric design, the complete set of tokens is processed by the lightweight decoder only, which significantly reduces the pre-training time. The encoder and decoder structures are shown in Figure 4.

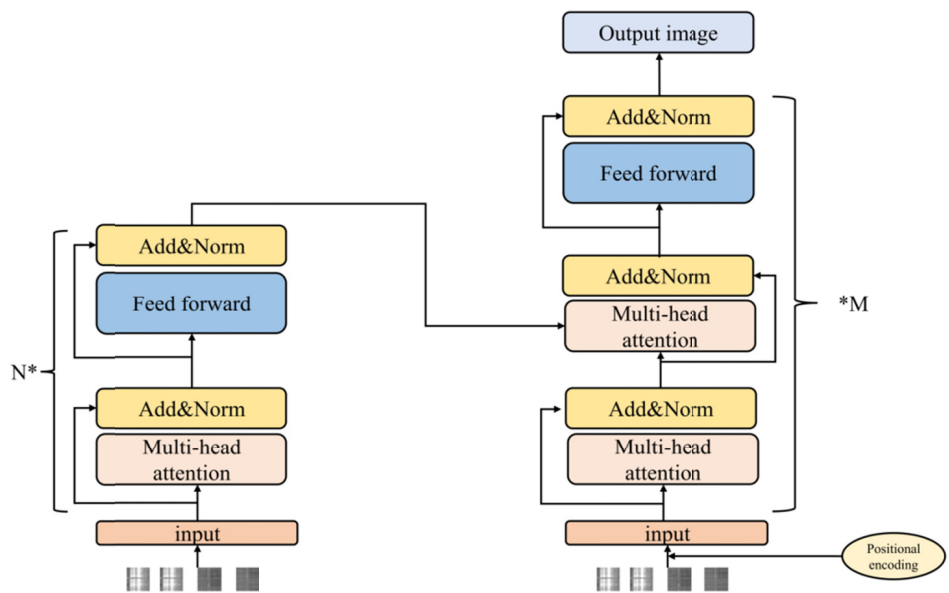

**Figure 4.** Encoder (**left**) and decoder (**right**) structures.

*3.4. Data Enhancement*

Because solar bursts are low-probability events, the solar radio burst spectra observed are not much. To improve the generalization ability and avoid the overfitting of the model, we augment the spectrum of solar radio bursts. We adopt a simple and data-independent augmentation method called mixup. Mixup can be implemented in only a few lines of code. A virtual training sample is constructed with minimal computational overhead. In an extensive evaluation, the results show that mixup improves the generalization error of the most advanced models in ImageNet, CIFAR, voice and tabular datasets. In addition, mixup helps to eliminate the memory of false labels, the sensitivity to confrontation samples and the instability of confrontation training. The formula for mixup data enhancement is as follows:

$$\tilde{x} = \lambda x_i + (1-\lambda) x_j \qquad (2)$$

$$\tilde{y} = \lambda y_i + (1 - \lambda) y_j \tag{3}$$

where $x_i$ and $x_j$ are two images randomly selected from the training set and $y_i$ and $y_j$ are their corresponding one-hot tags, respectively. Prior knowledge indicates that the linear interpolation of the sample images and the linear interpolation of the corresponding one-hot labels correspond. Mixup constructs a new sample and its one-hot label $\tilde{x}, \tilde{y}$ based on this prior knowledge. Among them, $\lambda$ is obtained by the data distribution $\beta(\alpha, \alpha)$, and $\alpha$ is a super parameter.

By adjusting the super parameter $\alpha$, we can adjust the proportion of interpolation between images. The research also shows that there is no good method to set $\alpha$ at present and the sensitivity to $\alpha$ is different in different datasets.

To further improve the generalization ability of the model and avoid overfitting, dropout is also introduced in this model. In deep learning, if the amount of data is small and the model is complex, the trained model will easily overfit. Dropout can alleviate the overfitting problem and achieve a regularization effect to a certain extent.

## 4. Experimental Dataset and Experimental Configuration

### 4.1. Experimental Dataset

A solar radio burst corresponds to a certain solar activity event, which is a low-probability event. The calibration data are relatively small and a large number of solar radio spectra are no-burst spectra. Therefore, there is an imbalance between the three types of samples.

Since the solar radio spectrum has two parts, left-handed and right-handed polarization, we separated the two parts so that the burst and calibration data can be expanded. The amount of solar radio data after amplification is shown in Table 1. The quantity of the three types of spectra is roughly in a balanced state, which alleviates the data imbalance problem. The experiment shows that the separation of these two parts has no effect on the results of solar radio spectrum classification.

After data enhancement, the total number of samples is 5519. We divided these randomly according to a proportion of approximately 8:2; 4415 labeled images were used as the training set and the remaining 1104 were used as the testing set. The specific number of samples in each category of the dataset is shown in Table 1.

**Table 1.** Number of solar radio spectrum samples.

| Type | Training Set | Testing Set | Total |
|---|---|---|---|
| Non-burst | 1648 | 412 | 2060 |
| Burst | 1476 | 369 | 1845 |
| Calibration | 1292 | 322 | 1614 |

### 4.2. Experimental Configuration and Evaluation Index

The software environment for our experiments is Windows 10, the programming platform is PyCharm, and the architecture is PyTorch. In the hardware device, the CPU is an Intel Core i7-10700k, the memory is 32 GB, and the GPU is a NVidia GeForce RTX 2080Ti. For pre-training, the batch size is 16, the image size is 224 × 224, the epoch is set to 300, the learning rate is $1.5 \times 10^{-4}$, the warmup learning rate is $10^{-6}$, the warmup is 30 epochs, and the weight decay is set to 0.05. In the fine-tuning stage, the epoch is set to 50, the learning rate is $10^{-3}$, the warmup learning rate is $10^{-6}$, the warmup is five epochs, and the other parameters are unchanged.

In practice, the burst class has greater research value and greater impact on daily life, so we mainly focus on the burst class. The burst class is defined as a positive class, and the other two classes are defined as negative classes. We define TP as the number of samples that are positive and correctly classified as positive, FP as the number of samples that are negative but wrongly classified as positive, TN as the number of samples that are negative and correctly classified as negative, and FN as the number of samples that are positive but classified as negative. The evaluation metrics used in the experiment are accuracy, precision, recall, specificity, and F-score.

Accuracy refers to the proportion of the number of accurate samples classified by all categories to the total number of samples. It is calculated as follows:

$$Accuracy = \frac{TP+TN}{TP+TN+FP+FN} \tag{4}$$

Precision refers to the proportion of the number of correctly predicted positive samples to the total number of predicted positive samples. It is calculated as follows:

$$Precision = \frac{TP}{TP+FP} \tag{5}$$

Recall refers to the proportion of correctly predicted positive samples relative to all actual positive samples. It is calculated as follows:

$$Recall = \frac{TP}{TP+FN} \tag{6}$$

Specificity refers to the proportion of correctly predicted negative cases relative to all actual negative cases. It is calculated as follows:

$$Specificity = \frac{TN}{TN+FP} \tag{7}$$

The balanced F-score is used for the overall evaluation of precision and recall. It is calculated as follows:

$$F_1 = 2 \times \frac{Precision \times Recall}{Precision + Recall} \tag{8}$$

## 5. Experimental Results and Discussion

### 5.1. Effect of Masking Rate on Classification Accuracy

When using the self-masking model structure for training, we divide the solar radio spectrum image into blocks and then mask these blocks randomly and send them to the network for training. We found that different masking rates affect training and classification accuracy. After using different masking rates for the experiments, the influence of masking rates on classification accuracy is shown in Figure 5.

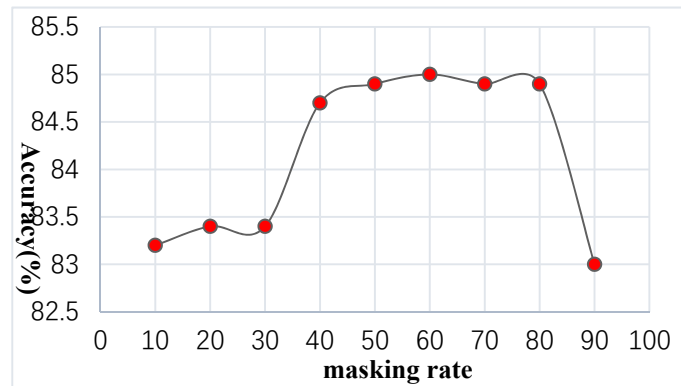

**Figure 5.** Influence of masking rate on classification accuracy.

In Figure 5, the experimental data show that a high masking rate corresponds to a high accuracy. When the masking rate is between 60% and 75%, the classification accuracy is the highest. The experimental results are counterintuitive. In natural language processing, the best masking rate of the BERT method is about 15%. In our results, the optimal masking rate is much higher. This may be because compared with text, images have more redundant information, while text contains more dense information.

Therefore, masking more information in the image can remove much redundant information unrelated to classification. This is more conducive to learning useful features and more essential information from the image, which is conducive to subsequent transfer learning. In this experiment, a random masking rate of 75% is finally selected and the masking effect is shown in Figure 6.

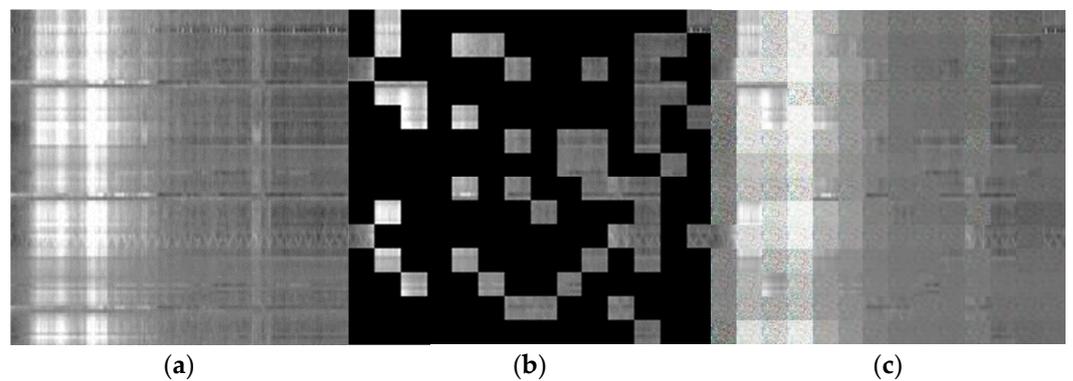

(**a**)            (**b**)            (**c**)

**Figure 6.** BERT masking experiments. (**a**) Original image; (**b**) after 75% masking; (**c**) restored images.

*5.2. Pre-Training with Transfer Learning*

After expanding the solar radio spectrum image dataset there are still few experimental data, therefore, the effect of training the mask model with these data, as shown in Figure 3, is poor. Therefore, we adopt the transfer learning training method. First, we train the mask model on the large dataset ImageNet [31]. At this stage, we do not need to use the tags in the ImageNet dataset. Then, we remove the encoder module and connect it to the fully connected layer and transfer it to the small dataset of the solar radio spectra for fine-tuning. An accuracy comparison using transfer learning is shown in Table 2.

As seen in Table 2, the transfer learning ability of the model is very good, which greatly improves the accuracy of the experiment. This is also because the Vision Transformer (ViT) structure requires a large amount of training data. Although the ViT model achieves good results, it is based on the use of a larger dataset than ImageNet. If only ImageNet is used, the results of the ViT model are not better than those of the CNN structure model. However, because the self-supervised method can learn more information about images than the supervised method, the information contained in the image itself is also far greater than the information contained in the label, therefore, the transfer learning effect of the self-supervised model is very good.

**Table 2.** Transfer learning effect.

| Method | Accuracy(%) |
|---|---|
| Training from the beginning | 70.60 |
| Transfer learning | 98.63 |

*5.3. Data Enhancement in Training*

In the process of ViT training, to further alleviate overfitting and accelerate convergence, we used many data enhancement methods. In our experiment, we tried color change, random erasure, mixup, cutmix, and their combination. The experimental results are shown in Table 3.

**Table 3.** Results of data enhancement experiments.

| | Mixing Method | | Accuracy (%) | Recall (%) |
|---|---|---|---|---|
| | Random Erasing | | 96.7 | 97.8 |
| | Color change | | 98.6 | 98.4 |
| | +Quadratic interpolation | +Random | 98.7 | 97.3 |
| Mixup | +Triple interpolation | +Batch | 98.1 | 97.8 |
| | +Quadratic interpolation | +Batch | 98.7 | 98.4 |
| | +Triple interpolation | +Random | 98.9 | 97.6 |
| Cutmix | +Batch | | 98.8 | 97.0 |
| | +Random | | 98.7 | 97.3 |
| Mixup + Cutmix | | | 99.3 | 98.4 |

Table 2 shows that based on the accuracy of 98.63% after transfer learning, mixup and cutmix improve the accuracy, the other two methods have no significant improvement effect on the final accuracy, and changing the color even makes the accuracy decrease slightly. Experiments were also conducted to determine how mixup generates virtual data. The experiments were divided into four groups. The accuracy of the model is higher with three interpolations plus the source of the same batch. Experiments were also conducted for the source of cutmix data, which were divided into two groups. The experimental results show that the results are better when the data come from the same batch. The final combination of mixup and cutmix yielded an accuracy of 99.3%, as shown in Table 3.

*5.4. Dropout*

We conducted experiments on different dropout methods, including dropout, DropPath, and DropAttention [32]. Dropout discards nodes in the network with a certain probability, DropPath discards paths in the network, and DropAttention discards the attention weight in the Transformer with a certain probability. The three methods can prevent overfitting, overcome network degradation and improve the network effect. The best results are shown in Table 4. In the three dropout experiments, DropPath is relatively good and the drop probability is 0.1.

**Table 4.** Results of dropout experiments.

| Method | Accuracy (%) | Recall (%) |
|---|---|---|
| Dropout | 98.8 | 97.8 |
| DropPath | 98.9 | 98.4 |
| DropAttention | 98.6 | 98.4 |

*5.5. Final Classification Results*

After integrating the above methods, the final results of the model and the change curve of its loss are shown in Figure 7. The final accuracy of the model is 99.5%. Compared with the results of some previous studies, it showed a good improvement effect and good migration results are achieved. Although the final restored image is relatively fuzzy, the restored image exhibits the characteristics of the solar radio burst images well. This shows that the network learns the characteristics of solar radio burst spectrum images well.

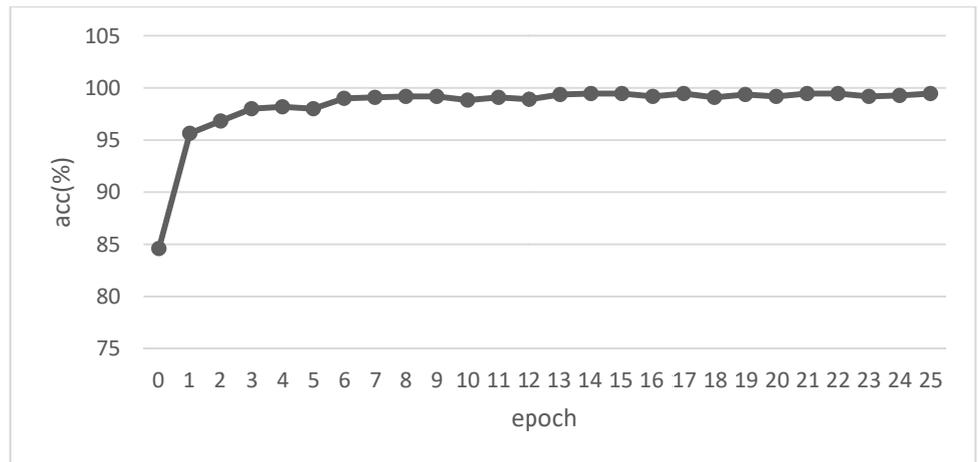

(a)

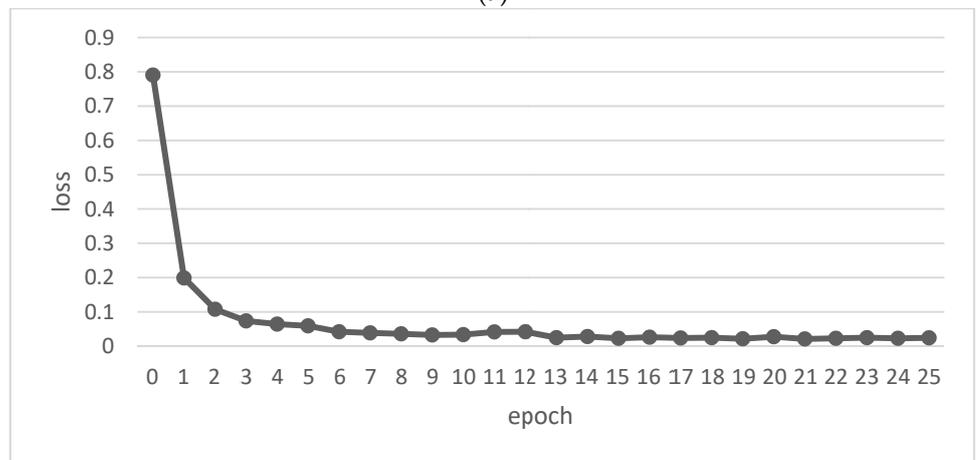

(b)

**Figure 7.** Model training results. (**a**) Accuracy curve, (**b**) Loss curve.

To measure the effectiveness of our model from various aspects, other indicators were considered in the experiment to evaluate our model. First, the confusion matrices of the three categories are calculated and the results are shown in Figure 8. Then, the confusion matrix is used to calculate the precision, recall and specificity of each category. In addition, the F1 score of each category is calculated. An F1 score of 1 is the best and 0 is the worst. The results are shown in Table 5.

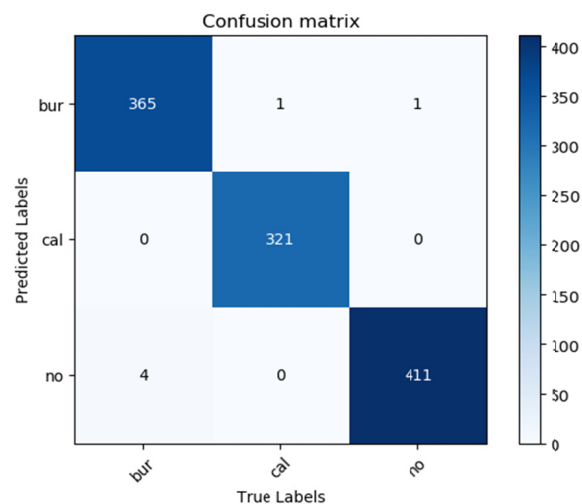

**Figure 8.** Confusion matrix.

**Table 5.** Experimental results of our model.

| Type | Precision (%) | Recall (%) | Specificity (%) | F1 Score |
|---|---|---|---|---|
| Burst | 99.5 | 98.9 | 99.7 | 0.992 |
| Non-burst | 99.0 | 99.8 | 99.4 | 0.994 |
| Calibration | 100 | 99.7 | 100 | 0.998 |

Our model is also compared with other models. The comparison model uses Vision Transformer, Swin Transformer, VGG, GoogLeNet, MoblieNet, ResNet and DenseNet as the core of the experimental network. For a fair comparison, other models are also migrated to the solar radio spectrum image for fine adjustment after pre-training. The results are shown in Table 6.

**Table 6.** Comparison experiments.

| Models | Accuracy (%) | Recall (%) |
|---|---|---|
| Vision Transformer | 94.3 | 97.5 |
| VGG16 | 96.0 | 98.4 |
| Swin Transformer | 99.0 | 99.1 |
| Resnet | 98.9 | 98.9 |
| MobileNet | 95.6 | 98.1 |
| GoogLeNet | 96.6 | 98.4 |
| DenseNet | 99.1 | 99.1 |
| **Ours** | **99.5** | **99.7** |

The self-supervision model can achieve the same accuracy as the current mainstream CNN and Transformer models. At the same time, compared with Vision Transformer, the accuracy of our model is obviously higher under the same transfer learning conditions. This shows that self-supervised learning is indeed more conducive to transfer learning than supervised learning.

For this application, we pay more attention to solar radio bursts that occurred with low probability. It is more critical to find all solar radio bursts as far as possible. Therefore, a high recall rate is important. Our method has achieved a 99.7% recall rate, which is better than other models.

## 6. Conclusions

In this paper, we propose a solar radio spectrum classification method based on self-supervised learning. By referring to the BERT method in natural language processing, it is improved for solar radio spectrum classification. After randomly masking the solar radio spectrum image, the method lets the model learn to restore the image to learn the image features. This method uses the image itself as a label to enable the network to learn more information, therefore, it is very suitable for transfer learning, thus addressing the issue of fewer solar radio spectrum image datasets. This method can also obtain a good result on a small dataset. Through experiments, an accuracy of 99.5% was achieved on the solar radio spectrum dataset. Compared with other models, our model achieves better experimental accuracy. However, our model has a larger number of parameters and requires more training time. Therefore, we need to continue to study how to reduce the scale of the model.

According to the spectral morphology of solar radio bursts, they can be divided into type I, II, III, IV, V and their associated fine structures. Different solar radio bursts correspond to different solar physical phenomena. Next, we will further subdivide and label the burst samples and use the fine-grained method to classify type I, II, III, IV, V bursts.

**Author Contributions:** S.L. proposed the network architecture design and the framework; J.C. and S.L. collected and preprocessed the datasets; S.L. performed the experiments; J.C. and S.L. analyzed and discussed the experimental data; S.L. and G.Y. wrote the article; H.Z. and C.T. revised the article and provided valuable advice for the experiments. All authors have read and agreed to the published version of the manuscript.

**Funding:** This research was funded by the Natural Science Foundation of China (Grant No. 12263008, 11941003), the MOST (Grant No. 2021YFA1600500), the Application and Foundation Project of Yunnan Province (Grant No. 202001BB050032), the Yunnan Provincial Department of Science and Technology-Yunnan University Joint Special Project for Double-Class Construction (Grant No. 202201BF070001-005), the Key R&D Projects in Yunnan Province (Grant No. 202202AD080004), the Open Project of CAS Key Laboratory of Solar Activity, the National Astronomical Observatories (Grant No. KLSA202115) and the Postgraduate Innovation Project of Yunnan University (Grant No. 2021Y269).

**Data Availability Statement:** The data are available at GitHub: https://github.com/filterbank/spectrumcls.

**Acknowledgments:** We would like to thank the anonymous reviewers and the editor-in-chief for their comments to improve the article. Thanks also to the data sharer. We thank all the people involved in the study.

**Conflicts of Interest:** The authors declare no conflicts of interest.